\documentclass{INTERSPEECH2023}

\interspeechcameraready 
\usepackage{xcolor}
\usepackage{balance}

\title{Transfer Learning for Personality Perception via Speech Emotion Recognition}
\name{Yuanchao Li, Peter Bell, Catherine Lai}
\address{Centre for Speech Technology Research, University of Edinburgh, UK}
\email{yuanchao.li@ed.ac.uk}

\begin{document}
\newcommand{\cfbox}[2]{%
    \colorlet{currentcolor}{.}%
    {\color{#1}%
    \fbox{\color{currentcolor}#2}}%
}

\maketitle

\begin{abstract}
Holistic perception of affective attributes is an important human perceptual ability. However, this ability is far from being realized in current affective computing, as not all of the attributes are well studied and their interrelationships are poorly understood. In this work, we investigate the relationship between two affective attributes: personality and emotion, from a transfer learning perspective. Specifically, we transfer Transformer-based and wav2vec2-based emotion recognition models to perceive personality from speech across corpora. Compared with previous studies, our results show that transferring emotion recognition is effective for personality perception. Moreoever, this allows for better use and exploration of small personality corpora. We also provide novel findings on the relationship between personality and emotion that will aid future research on holistic affect recognition.
\end{abstract}
\noindent\textbf{Index Terms}: personality perception, emotion recognition, Transformer, wav2vec2, transfer learning, data augmentation

\section{Introduction}

Humans have the ability to perceive affective aspects of speech holistically. However, in affective computing, different attributes are usually recognized in isolation within one relatively narrow corpus, resulting in their relationships being understudied \cite{zhang2019holistic}. To address this issue, multitask learning has been employed to jointly recognize multiple attributes using a shared model. For instance, researchers have successfully recognized emotion in conjunction with gender, speaker, or personality \cite{li2019improved,fu2021end,li2021multitask}. However, joint training may be hindered by one specific task dominating the training process, negatively impacting the overall performance due to inductive transfer \cite{zhang2019holistic,li2019improved}. Therefore, investigating the relatedness of tasks is of utmost importance. In this work, we propose to study personality and its relationship with emotion using transfer learning. Compared to emotion recognition, which has been extensively researched over the past decade, personality perception attracts much less attention. Therefore, we aim to provide new insights for personality research and to investigate the relationship between personality and emotion towards the goal of holistic affect recognition from speech.

Personality is a psychological construct aimed at explaining human behaviors in terms of a few stable and measurable individual characteristics, which are kept relatively steady over time \cite{vinciarelli2014survey}. The most widely used theory for describing personality is the Big-Five traits \textit{OCEAN}: \textit{OPenness (OP)}, \textit{COnscientiousness (CO)}, \textit{EXtraversion (EX)}, \textit{AGreeableness (AG)}, \textit{NEuroticism (NE)}. These traits can substantially affect human perception and result in various behaviors, such as facial and vocal expressions, eye gaze, body postures and gestures \cite{vinciarelli2009social}. Thus, similar to other affective computing topics, studies on personality computing also focus on recognition and synthesis tasks. The recognition task aims at perceiving personality traits from multimodal features. For example, \cite{okada2015personality} used high-level features obtained from body and head motion, speaking activity, along with low-level features extracted from audio, to predict the five personality traits. The synthesis task tries to generate behaviors to express artificial personalities for human-computer interaction. For instance, \cite{yamamoto2021character} controlled the amount of utterances, backchannels and fillers, and switching pause length for a robot to express extraversion, emotional instability and politeness. With the recognition and synthesis functions, systems can be designed to have behaviors similar to those of human beings in particular scenarios such as job interview and health care.

Although personality can be generally understood from speech \cite{Polzehl2015PersonalityIS}, research has focused on use of multimodal signals rather than only speech \cite{okada2015personality}. Despite providing potentially important affective information, studies on personality corpora usually have fewer citations than those about emotion corpora from the past decade. One of the reasons for these is that there are very few large datasets designed for speech research for personality, which makes it harder to apply deep learning methods. However, as many issues remain unresolved or undiscovered in speech-based personality analysis, we argue that small corpora are still worth exploring. To this end, in this work, we investigate several methods which are particularly suitable for small personality corpora, and show how they can produce novel findings, even with only a small amount of data labeled. We demonstrate that \textbf{1)} transferring a wav2vec2-based emotion recognition model to personality perception is feasible; \textbf{2)} the speech representations for personality perception and emotion recognition are highly transferable at the phonetic level; \textbf{3)} personality is likely to be more related to arousal than to valence; \textbf{4)} \textit{CO} is less related to emotion than other personality traits; \textbf{5)} some data augmentation techniques for speech recognition may not be suitable for personality perception.

\section{Related Work}

\subsection{Personality Perception Using SPC}
The SSPNet Speaker Personality Corpus (SPC) \cite{mohammadi2012automatic} was one of the first corpora for personality computing and was one of the benchmark corpora analyzed in the INTERSPEECH 2012 speaker trait challenge \cite{schuller2012interspeech}. It consists of 640 speech clips in French for a total of 322 subjects.  Most speech clips last 10 seconds, and each was scored by 11 annotators in terms of the Big-Five traits using the BFI-10 questionnaire. Information on speaker gender, speaker status (journalist or non-journalist), and speaker ID were also attached to each speech clip. We processed the labels in the same way as the Speaker Trait Challenge \cite{schuller2012interspeech}: each clip was labeled as above average $S$ for a specific trait $S \in \{EX, AG, CO, NE, OP\}$ if at least six judges (the majority) gave it a score higher than their average for the same trait; otherwise, it was labeled $\overline{S}$.

In the speaker trait challenge, a set of acoustic features, such as energy, spectral and voicing related low-level descriptors as well as their functionals, were selected as the baseline for subsequent personality perception works. Several novel methods and findings come out of the analysis of SPC. For example, \cite{carbonneau2017feature} proposed a feature learning method which learns a dictionary of discriminant features from patches extracted in spectrograms to simplify the feature extraction process. \cite{gilpin2018perception} found that predicting \textit{EX} using classifiers trained only on the male, female, or journalist subgroups resulted in higher accuracies than all of these groups combined. \cite{fayet2017big} demonstrated features based on psychological information can bring more information to personality perception than audio features only. Despite this progress, SPC's potential has not been fully exploited due to its small size. Thus, we propose a method--transferring emotion recognition to personality perception, hoping to obtain new insights into the relationship between personality and emotion.

\subsection{Relationship Between Personality and Emotion}
In order to use transfer learning, we need to ensure that personality and emotion have mutual effects. Previously, \cite{barford2016openness} conducted a theoretically grounded investigation of personality and emotion, demonstrating that \textit{NE} predicted higher scores of negative emotions and lower scores of positive emotions, whereas \textit{AG} predicted the opposite. In \cite{mohammadi2020multi}, positive emotions were found to be positively correlated with \textit{EX, AG, CO}, and \textit{OP}. On the other hand, \textit{NE} was positively correlated with negative emotions and negatively correlated with positive emotions. Moreover, deep learning tasks have been shown to benefit from the relationship between personality and emotion. \cite{li2019attention} proposed integrating the target speaker’s personality embeddings into the learning of an attention-based network to improve Speech Emotion Recognition (SER) performance. \cite{li2021multitask} used multitask learning to detect emotion and personality, achieving competitive results across multiple datasets. These works demonstrate the existence of a link between personality and emotion, and inspire us to continue to study personality, taking into account its relationship with emotion.

\section{Methodology}

\begin{figure*}[t]
  \centering
  \includegraphics[width=0.9\textwidth]{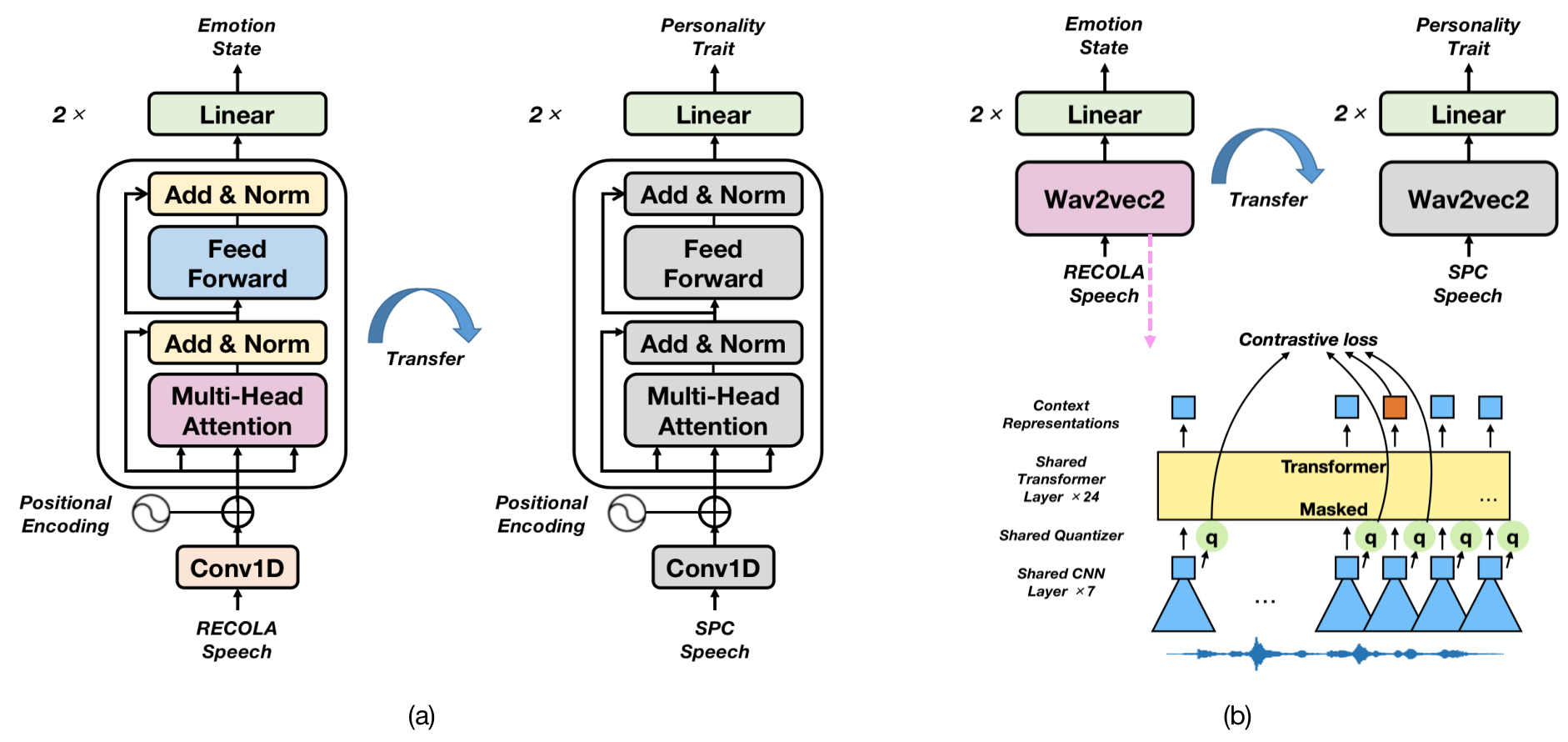}
  \caption{Proposed transfer learning models. (a): Transformer-based; (b): wav2vec2-based. Colored: trainable. Gray: frozen.}
  \label{fig:model}
\vspace{-10pt}
\end{figure*}

We propose to use transfer learning to pre-train two SER models, which are based on a Transformer architecture and a self-supervised architecture--wav2vec2, respectively. In Fig~\ref{fig:model}, (a) is the Transformer-based model, while (b) shows the wav2vec2-based model with the depiction of a wav2vec2 architecture. To make comparisons for obtaining more findings, we also used data augmentation, which is another popular method for solving data scarcity problems.

\subsection{Transfer Learning}
\noindent\textbf{RECOLA dataset.}
We used the RECOLA dataset \cite{ringeval2013introducing}  to build the emotion recognition model. RECOLA includes audiovisual and physiological recordings of dyadic interactions from 27 French-speaking subjects, and was annotated continuously for arousal and valence dimensions by six gender-balanced French-speaking annotators. Data from 23 subjects is publicly available and only its audio files were used in this work. To make the labeling consistent, the same practice as SPC was used for the arousal and valence dimensions. Furthermore, to make pre-training on RECOLA and fine-tuning and testing on SPC consistent, we cut each audio file of RECOLA (5min) into 30 clips (10s). We also used 5s clips to generate more training data, but the difference in the experimental results was negligible.

\noindent\textbf{Transformer-based model.}
First, the speech waveform was fed into a Convolutional Neural Network (CNN) whose window size is 3 and stride is 2. The positional encoding vectors are of the same dimension as the CNN output for the summation to inject the positional information. Then, we used the encoder part of the Transformer (but only one layer due to the small size of RECOLA) which is composed of two sub-layers: Multi-Head Attention (MHA) and Feed Forward Network (FFN). For the MHA, we used eight heads to receive eight different linearly projected versions of the query, key, and value, to produce eight outputs that are used to generate a final result.

The FFN consists of two linear transformations with ReLU activation in between. In addition, each of the two sub-layers has a residual connection and is also succeeded by a normalization layer, which normalizes the sum computed between the sub-layer input and the output generated by the sub-layer itself. Following the Transformer encoder, two linear layers with softmax activation were implemented for the classification tasks. The full model was trainable during the pre-training on RECOLA and frozen, except for the linear layers, during fine-tuning on SPC. 

\noindent\textbf{Wav2vec2-based model.}
Wav2vec2 is a self-supervised learning framework that discriminates positive (true) frame samples from negative ones while minimizing contrastive prediction loss. Among a set of wav2vec2 models, wav2vec2-xlsr-53 is a large cross-lingual model, which learns speech representations by pre-training a single model from the raw waveform of speech in 53 languages \cite{conneau2020unsupervised}. It consists of three major components: a CNN-based local encoder that extracts a sequence of embeddings from raw audio as latent representation; a Transformer network for context representation; and a shared quantizer for discretization across multiple languages. In this work, we adopted the \textit{wav2vec2-large-xlsr-53-french} model which was fine-tuned on French using the training and validation splits of CommonVoice. As with the Transformer-based model, we trained the full model and froze wav2vec2 for fine-tuning. We only trained the last half of layers in wav2vec2 as prior work has demonstrated that the ability to capture acoustic information in the first half layers is barely affected by further training \cite{li2023exploration}. That work also found that if the SER result of using the middle layer output clearly outperforms using the average of all layer outputs, then it means SER on this corpus depends largely or only on acoustics. Thus, we compared both the middle layer and average layer outputs to investigate speech-based personality perception. Note that there are other self-supervised speech models that can also be used, but since our purpose is to find the relationship between personality and emotion rather than to explore the models, we choose to use a French version that is suitable for both RECOLA and SPC.

\subsection{Data Augmentation}
Data augmentation is a widely used method for solving data scarcity problems in speech research, such as speech recognition \cite{park2019specaugment}. To our best knowledge, only one study has used this method for personality perception. However, it did not yield good results or explain why this was the case \cite{zaferani2021automatic}. It may be because they used several masking methods separately, resulting in little difference between the newly generated data and the original data. Here, we further explore whether data augmentation works or why it fails for personality perception.

To augment the data, we applied frequency masking and time masking from SpecAugment \cite{park2019specaugment} and their combinations: frequency masking followed by time masking and vice versa, to make the size of SPC four times larger for a fair comparison with the transfer learning methods. Then, for classification, we constructed the a CNN and two linear layers with a softmax, the same as the Transformer-based model. To avoid overfitting, dropout with a probability of 0.5 was implemented after every layer.

\section{Experimental Evaluation}
\subsection{Implementation}

The model was built using Pytorch and optimized using the AdamW method. The learning rate and weight decay were set at 1e-4 and 1e-5, respectively. To minimize the overall loss for the binary classification, cross-entropy was used. For the transfer learning method, the full model was trained on RECOLA ten times from 10 to 100 epochs (increased by 10 epochs each time) using arousal and valence labels, respectively. This is to compare the effects of the amount of pre-training on SER to investigate the connection between different personality traits and emotion dimensions. If more training epochs improves personality perception performance, then we have evidence that the personality trait is closer to that emotion dimension. After each time, 5-fold (80\% for fine-tuning, 20\% to testing) cross-validation was conducted on SPC for 100 epochs. For the data augmentation method, the model was trained for 100 epochs using 5-fold cross-validation on SPC only. We evaluated the performance using the Unweighted Average Recall (UAR).

\subsection{Results and Discussion}

We present our results and the comparison with major previous studies in Table~\ref{tab:result}. As in \cite{mohammadi2012automatic}, Logistic Regression (LR) and SVM were used for classification from energy, pitch, first two formants, and length of voiced and unvoiced segments, as well as their functionals. In the INTERSPEECH 2012 speaker trait challenge \cite{schuller2012interspeech}, Random Forest (RF) and SVM classifiers were trained with 6125-dimensional low-level features. \cite{carbonneau2017feature} compared the performance using four types of features: GeMAPS, eGeMAPS, outputs from a Sparse Auto-Encoder (SAE) and a stacked SAE. They also proposed a feature learning method which learns a dictionary of discriminant features from patches extracted in the spectrogram representations for encoding the speech input. An SVM was used as the classifier for all five feature types. \cite{solera2017semi} used weak supervision to iteratively refine the initial model trained on labeled data using the unlabeled data based on knowledge-based features that exploit expert knowledge of acoustic-prosodic cues. \cite{zaferani2021} employed an autoencoder to extract features and used an adaptive neuro-fuzzy inference system to model the uncertainty in speech.

From the comparison results with the literature, it can be noted that transfer learning-based methods achieve comparable results to those in the literature. In particular, \textbf{1)} by comparing the best scores of the transfer learning methods (in \textbf{bold}) with the literature, we can see that they achieve very competitive results for all the traits except \textit{CO}. This may be because Conscientiousness has less similarity with arousal and valence. \textbf{2)} The wav2vec2-based (mid.) model trained with arousal labels achieves better average UAR score (in\colorbox{lime}{green}) than half of the baselines and their average (in\colorbox{lightgray}{gray}). This shows that non-linguistic speech representations of arousal are highly transferable to those of personality traits in general. As the middle layer of wav2vec2 contains mostly phonetic-level features \cite{pasad2021layer}, we believe that the joint recognition of personality and emotion can benefit the most from using these features.

\begin{table}[t]
\centering
  \caption{Performance comparison in terms of UAR (\%).}
  \label{tab:result}
\scalebox{0.84}{
\begin{tabular}{lcccccc}
\toprule
\textbf{Method} & \textbf{\textit{EX}} & \textbf{\textit{AG}} & \textbf{\textit{CO}} & \textbf{\textit{NE}} & \textbf{\textit{OP}} & \textbf{Avg.} \\ \midrule
\multicolumn{7}{c}{\textit{Literature}} \\ \midrule
\cite{mohammadi2012automatic}-\textit{LR} & 72.4 & 55.7 & 69.6 & 67.4 & 56.1 & \colorbox{lightgray}{64.2} \\
\cite{mohammadi2012automatic}-\textit{SVM} & 74.3 & 57.4 & 68.0 & 65.5 & 57.7 & \colorbox{lightgray}{64.6} \\
\cite{schuller2012interspeech}-\textit{RF} & 77.5 & 60.1 & 69.0 & 68.2 & 52.9 & \colorbox{lightgray}{65.5} \\
\cite{schuller2012interspeech}-\textit{SVM} & 74.5 & 62.2 & 69.2 & 69.0 & 58.7 & 66.7 \\
\cite{carbonneau2017feature}-\textit{GeMAPS} & 74.9 & 61.9 & 72.2 & 68.9 & 56.3 & 66.8 \\
\cite{carbonneau2017feature}-\textit{eGeMAPS} & 75.1 & 62.0 & 72.5 & 66.6 & 53.7 & 66.0 \\
\cite{carbonneau2017feature}-\textit{SAE 1 layer} & 69.2 & 62.0 & 64.3 & 65.8 & 57.1 & \colorbox{lightgray}{63.7} \\
\cite{carbonneau2017feature}-\textit{SAE 2 layers} & 69.0 & 60.3 & 63.6 & 61.9 & 57.3 & \colorbox{lightgray}{62.4} \\
\cite{carbonneau2017feature}-\textit{Feature learning} & 75.2 & 64.9 & 68.3 & 70.8 & 56.3 & 67.1 \\
\cite{solera2017semi}-\textit{Weak supervision} & 68.5 & 57.0 & 70.8 & 66.2 & 54.8 & \colorbox{lightgray}{63.5} \\
\cite{zaferani2021}-\textit{Autoencoder} & 62.9 & {71.8} & 69.1 & 70.9 & 68.9 & 68.7 \\
\textit{Average} & 72.1 & 61.4 & 68.9 & 67.4 & 57.3 & \colorbox{lightgray}{65.4} \\ \midrule
\multicolumn{7}{c}{\textit{Ours}} \\ \midrule
\multicolumn{7}{l}{Transformer-based} \\
\ \ \ \ \textit{Arousal}  & 68.9 & 62.2 & 60.5 & 63.8 & 63.2 & 63.7 \\
\ \ \ \ \textit{Valence}  & 69.6 & 61.0 & 58.8 & 64.6 & 61.9 & 63.2 \\ 
\multicolumn{7}{l}{Wav2vec2-based (mid.)} \\
\ \ \ \ \textit{Arousal}  & \textbf{74.6} & 61.8 & \textbf{62.3} & 65.8 & \textbf{63.8} & \colorbox{lime}{65.7} \\
\ \ \ \ \textit{Valence}  & 72.9 & 62.7 & 59.4 & 60.2 & \textbf{63.8} & 63.8 \\ 
\multicolumn{7}{l}{Wav2vec2-based (avg.)}  \\
\ \ \ \ \textit{Arousal}  & 70.2 & 62.7 & 55.0 & \textbf{69.9} & 59.2 & 63.4 \\
\ \ \ \ \textit{Valence}  & 68.6 & \textbf{63.5} & 57.7 & 56.5 & 62.4 & 61.7 \\ 
{Data augmentation} & 62.6 & 59.8 & 61.6 & 58.8 & 54.5 & 59.5 \\ \bottomrule
\end{tabular}}
\end{table}

By looking only at our results, we see that \textbf{1)} the wav2vec2-based model using the middle layer output outperforms other models and is better than the one using the average layer output. This phenomenon is also because of the characteristics of wav2vec2: the first half layers of wav2vec2 are responsible for encoding acoustic information while the second half encodes linguistic information \cite{pasad2021layer,li2023exploration}. Since the training was performed on RECOLA, which has different linguistic content, such a difference resulted in the second half layers having difficulty capturing linguistic information, which drags down the overall performance. \textbf{2)} Wav2vec2-based (mid.) is better than Transformer-based, which could be for two reasons. First, the emotion information learned by self-supervised contrastive loss is transferable to personality. Second, the middle layer of wav2vec2 captures stronger acoustic representations due to the sophisticated architecture. Note that although the wav2vec2 contains a deeper Transformer, it does not bring a significant overall improvement, indicating that there exists a performance bottleneck in transfer learning. \textbf{3)} The performance of wav2vec2-based (avg.) is worse than the Transformer-based, further suggesting that linguistic information is one of the limitations and concerns of using wav2vec2 in transfer learning.

Beyond this, models trained using arousal labels have better overall performance than those using valence labels. During the experiments we noticed that the more pre-training (60 to 100 epochs) the better the results when using arousal, but the opposite is true when using valence (10 to 50 epochs). These facts indicate that personality is closer to arousal as they both largely rely on acoustic information (how you say), while valence depends more on linguistic information (what you say) \cite{li2019expressing}. In contrast, the data augmentation method does not achieve satisfactory performance and is worse than transfer learning methods. The overfitting problem emerged very quickly during the model training, which demonstrates that some data augmentation techniques for speech recognition (e.g., SpecAugment) may not be suitable for personality perception. It could be because unlike speech recognition, personality perception is little affected by speech masking, which complies with the fact that personality is kept stable and steady over time \cite{vinciarelli2014survey}.

Last but not least, since we use transfer learning from emotion recognition for personality perception, it is inevitable that we cannot show a significant performance gain compared to the literature. As the purpose of this work is to study transferability of emotion based speech representations to a small personality corpus, we do not expect to achieve state-of-the-art performance. Note that some previous work reported their results in accuracy rather than UAR, so we cannot include them for performance comparison. For example, \cite{su2017personality} used multiresolution analysis based on wavelet transform to separate signals according to high and low frequency before feature extraction. In \cite{liu2020speech}, the variance of unknown differences in judgment’s perception of modified speech targets with hierarchical clustering was reduced. Before feature extraction, three filters were applied to the audio clips to avoid the uncertain effects of noise, silence, and pitch. These works demonstrate the usefulness of pre-processing the speech data before feature extraction. We expect this may also benefit transfer learning for personality perception.

\begin{table}[t]
\centering
  \caption{Correlation between every personality trait pair.}
  \label{tab:pcc}
\scalebox{0.92}{
\begin{tabular}{lcc}
\toprule
\textbf{Trait Pair} & \textbf{Phi (2-scale)} & \textbf{Pearson (5-scale)} \\ \hline
(\textit{EX}, \textit{AG}) & 0.03 & 0.30 \\ 
(\textit{EX}, \textit{CO}) & 0.02 & \textbf{0.50}  \\ 
(\textit{EX}, \textit{NE}) & 0.01 & 0.32 \\ 
(\textit{EX}, \textit{OP}) & 0.00 & \textbf{0.47} \\ 
(\textit{AG}, \textit{CO}) & 0.07 & 0.12 \\ 
(\textit{AG}, \textit{NE}) & 0.07 & \textbf{0.57} \\
(\textit{AG}, \textit{OP}) & 0.02 & 0.04 \\ 
(\textit{CO}, \textit{NE}) & 0.04 & 0.00 \\ 
(\textit{CO}, \textit{OP}) & 0.09 & \textbf{0.44} \\ 
(\textit{NE}, \textit{OP}) & 0.02 & 0.10 \\
\bottomrule
\end{tabular}}
\vspace{-10pt}
\end{table}

\subsection{An Attempt at Joint Perception of Trait Pairs}
As inspired by the joint recognition of arousal and valence on RECOLA \cite{kopru2020multimodal}, we also tried using multitask learning together with the transfer learning method. We replaced the last linear layer in Fig~\ref{fig:model} by a multitask head with two parallel linear layers. We aimed to predict arousal and valence jointly during training and to predict every pair of personality traits during fine-tuning and testing. However, the results did not change much from Table~\ref{tab:result} and sometimes only one trait was well perceived. To figure out the reason, we calculated the correlation coefficient (in absolute value) between every trait pair (Phi for 2-scale and Pearson for 5-scale). From Table~\ref{tab:pcc}, we can see that the correlation between every trait pair is blurred using the 2-scale labeling (as described in Sec.~2.1), which explains why the multitask head did not work here. As a comparison, the correlations are clear (in \textbf{bold}) in some trait pairs if using the original 5-scale labeling. We believe that the proper use of multitask learning and transfer learning together could yield better results and novel personality-emotion relationship findings in the future.

\section{Conclusions}
In this paper, we use transfer learning for personality perception via speech emotion recognition models. We obtain several findings on personality that have not been explored in previous research: \textbf{1)} transferring emotion-based speech representations to personality is useful as a connection exists between personality and emotion; \textbf{2)} phonetic-level features contain the most relatedness between personality and emotion (as opposed to linguistic features); \textbf{3)} models trained using arousal labels generally have better performance than those using valence labels; \textbf{4)} different personality traits have different patterns of relationships with emotion dimensions; \textbf{5)} transfer learning performs better than the type of augmentation provided by SpecAugment for personality perception. This work shows that small corpora are worth exploring in the deep learning area and can provide new knowledge in affective computing. This advances our long-term goal is to bridge the gap between different tasks to better understand holistic affect recognition from speech.

\balance
\bibliographystyle{IEEEtran}
\bibliography{mybib}

\begin{thebibliography}{10}
\providecommand{\url}[1]{#1}
\csname url@samestyle\endcsname
\providecommand{\newblock}{\relax}
\providecommand{\bibinfo}[2]{#2}
\providecommand{\BIBentrySTDinterwordspacing}{\spaceskip=0pt\relax}
\providecommand{\BIBentryALTinterwordstretchfactor}{4}
\providecommand{\BIBentryALTinterwordspacing}{\spaceskip=\fontdimen2\font plus
\BIBentryALTinterwordstretchfactor\fontdimen3\font minus
  \fontdimen4\font\relax}
\providecommand{\BIBforeignlanguage}[2]{{%
\expandafter\ifx\csname l@#1\endcsname\relax
\typeout{** WARNING: IEEEtran.bst: No hyphenation pattern has been}%
\typeout{** loaded for the language `#1'. Using the pattern for}%
\typeout{** the default language instead.}%
\else
\language=\csname l@#1\endcsname
\fi
#2}}
\providecommand{\BIBdecl}{\relax}
\BIBdecl

\bibitem{zhang2019holistic}
Y.~Zhang, F.~Weninger, B.~Schuller, and R.~W. Picard, ``Holistic affect
  recognition using panda: paralinguistic non-metric dimensional analysis,''
  \emph{IEEE Transactions on Affective Computing}, vol.~13, no.~2, pp.
  769--780, 2019.

\bibitem{li2019improved}
Y.~Li, T.~Zhao, and T.~Kawahara, ``Improved end-to-end speech emotion
  recognition using self attention mechanism and multitask learning.'' in
  \emph{Interspeech}, 2019, pp. 2803--2807.

\bibitem{fu2021end}
C.~Fu, C.~Liu, C.~T. Ishi, and H.~Ishiguro, ``An end-to-end multitask learning
  model to improve speech emotion recognition,'' in \emph{2020 28th European
  Signal Processing Conference (EUSIPCO)}.\hskip 1em plus 0.5em minus
  0.4em\relax IEEE, 2021, pp. 1--5.

\bibitem{li2021multitask}
Y.~Li, A.~Kazameini, Y.~Mehta, and E.~Cambria, ``Multitask learning for emotion
  and personality detection,'' \emph{arXiv preprint arXiv:2101.02346}, 2021.

\bibitem{vinciarelli2014survey}
A.~Vinciarelli and G.~Mohammadi, ``A survey of personality computing,''
  \emph{IEEE Transactions on Affective Computing}, vol.~5, no.~3, pp. 273--291,
  2014.

\bibitem{vinciarelli2009social}
A.~Vinciarelli, M.~Pantic, and H.~Bourlard, ``Social signal processing: Survey
  of an emerging domain,'' \emph{Image and vision computing}, vol.~27, no.~12,
  pp. 1743--1759, 2009.

\bibitem{okada2015personality}
S.~Okada, O.~Aran, and D.~Gatica-Perez, ``Personality trait classification via
  co-occurrent multiparty multimodal event discovery,'' in \emph{Proceedings of
  the 2015 ACM on International Conference on Multimodal Interaction}, 2015,
  pp. 15--22.

\bibitem{yamamoto2021character}
K.~Yamamoto, K.~Inoue, S.~Nakamura, K.~Takanashi, and T.~Kawahara, ``A
  character expression model affecting spoken dialogue behaviors,'' in
  \emph{Conversational Dialogue Systems for the Next Decade}.\hskip 1em plus
  0.5em minus 0.4em\relax Springer, 2021, pp. 3--13.

\bibitem{Polzehl2015PersonalityIS}
T.~Polzehl, ``Personality in speech,'' in \emph{Assessment and automatic
  classification}.\hskip 1em plus 0.5em minus 0.4em\relax Springer, 2015.

\bibitem{mohammadi2012automatic}
G.~Mohammadi and A.~Vinciarelli, ``Automatic personality perception: Prediction
  of trait attribution based on prosodic features,'' \emph{IEEE Transactions on
  Affective Computing}, vol.~3, no.~3, pp. 273--284, 2012.

\bibitem{schuller2012interspeech}
B.~Schuller, S.~Steidl, A.~Batliner, E.~N{\"o}th, A.~Vinciarelli, F.~Burkhardt,
  R.~Van~Son, F.~Weninger, F.~Eyben, T.~Bocklet \emph{et~al.}, ``The
  interspeech 2012 speaker trait challenge,'' in \emph{Interspeech}, 2012.

\bibitem{carbonneau2017feature}
M.-A. Carbonneau, E.~Granger, Y.~Attabi, and G.~Gagnon, ``Feature learning from
  spectrograms for assessment of personality traits,'' \emph{IEEE Transactions
  on Affective Computing}, vol.~11, no.~1, pp. 25--31, 2017.

\bibitem{gilpin2018perception}
L.~H. Gilpin, D.~M. Olson, and T.~Alrashed, ``Perception of speaker personality
  traits using speech signals,'' in \emph{Extended Abstracts of the 2018 CHI
  Conference on Human Factors in Computing Systems}, 2018, pp. 1--6.

\bibitem{fayet2017big}
C.~Fayet, A.~Delhay, D.~Lolive, and P.-F. Marteau, ``Big five vs. prosodic
  features as cues to detect abnormality in sspnet-personality corpus.'' in
  \emph{Interspeech}, 2017, pp. 3281--3285.

\bibitem{barford2016openness}
K.~A. Barford and L.~D. Smillie, ``Openness and other big five traits in
  relation to dispositional mixed emotions,'' \emph{Personality and individual
  differences}, vol. 102, pp. 118--122, 2016.

\bibitem{mohammadi2020multi}
G.~Mohammadi and P.~Vuilleumier, ``A multi-componential approach to emotion
  recognition and the effect of personality,'' \emph{IEEE Transactions on
  Affective Computing}, 2020.

\bibitem{li2019attention}
J.-L. Li and C.-C. Lee, ``Attention learning with retrievable acoustic
  embedding of personality for emotion recognition,'' in \emph{2019 8th
  International Conference on Affective Computing and Intelligent Interaction
  (ACII)}.\hskip 1em plus 0.5em minus 0.4em\relax IEEE, 2019, pp. 171--177.

\bibitem{ringeval2013introducing}
F.~Ringeval, A.~Sonderegger, J.~Sauer, and D.~Lalanne, ``Introducing the recola
  multimodal corpus of remote collaborative and affective interactions,'' in
  \emph{2013 10th IEEE international conference and workshops on automatic face
  and gesture recognition (FG)}.\hskip 1em plus 0.5em minus 0.4em\relax IEEE,
  2013, pp. 1--8.

\bibitem{conneau2020unsupervised}
A.~Conneau, A.~Baevski, R.~Collobert, A.~Mohamed, and M.~Auli, ``Unsupervised
  cross-lingual representation learning for speech recognition,'' \emph{arXiv
  preprint arXiv:2006.13979}, 2020.

\bibitem{li2023exploration}
Y.~Li, Y.~Mohamied, P.~Bell, and C.~Lai, ``Exploration of a self-supervised
  speech model: A study on emotional corpora,'' in \emph{2022 IEEE Spoken
  Language Technology Workshop (SLT)}.\hskip 1em plus 0.5em minus 0.4em\relax
  IEEE, 2023, pp. 868--875.

\bibitem{park2019specaugment}
D.~S. Park, W.~Chan, Y.~Zhang, C.-C. Chiu, B.~Zoph, E.~D. Cubuk, and Q.~V. Le,
  ``Specaugment: A simple data augmentation method for automatic speech
  recognition,'' \emph{arXiv preprint arXiv:1904.08779}, 2019.

\bibitem{zaferani2021automatic}
E.~J. Zaferani, M.~Teshnehlab, and M.~Vali, ``Automatic personality traits
  perception using asymmetric auto-encoder,'' \emph{IEEE Access}, vol.~9, pp.
  68\,595--68\,608, 2021.

\bibitem{solera2017semi}
R.~Solera-Ure{\~n}a, H.~Moniz, F.~Batista, R.~Cabarr{\~a}o, A.~Pompili,
  R.~Astudillo, J.~Campos, A.~Paiva, and I.~Trancoso, ``A semi-supervised
  learning approach for acoustic-prosodic personality perception in
  under-resourced domains,'' in \emph{18th Annual Conference of the
  International Speech Communication Association, INTERSPEECH 2017}.\hskip 1em
  plus 0.5em minus 0.4em\relax International Speech Communication Association,
  2017, pp. 929--933.

\bibitem{zaferani2021}
E.~J. Zaferani, M.~\vspace{0mm} Teshnehlab, and M.~Vali, ``Automatic
  personality perception using autoencoder and hierarchical fuzzy
  classification,'' in \emph{2021 26th International Computer Conference,
  Computer Society of Iran (CSICC)}.\hskip 1em plus 0.5em minus 0.4em\relax
  IEEE, 2021, pp. 1--7.

\bibitem{pasad2021layer}
A.~Pasad, J.-C. Chou, and K.~Livescu, ``Layer-wise analysis of a
  self-supervised speech representation model,'' in \emph{2021 IEEE Automatic
  Speech Recognition and Understanding Workshop (ASRU)}.\hskip 1em plus 0.5em
  minus 0.4em\relax IEEE, 2021, pp. 914--921.

\bibitem{li2019expressing}
Y.~Li, C.~T. Ishi, K.~Inoue, S.~Nakamura, and T.~Kawahara, ``Expressing
  reactive emotion based on multimodal emotion recognition for natural
  conversation in human--robot interaction,'' \emph{Advanced Robotics},
  vol.~33, no.~20, pp. 1030--1041, 2019.

\bibitem{su2017personality}
M.-H. Su, C.-H. Wu, K.-Y. Huang, Q.-B. Hong, and H.-M. Wang, ``Personality
  trait perception from speech signals using multiresolution analysis and
  convolutional neural networks,'' in \emph{2017 Asia-Pacific Signal and
  Information Processing Association Annual Summit and Conference (APSIPA
  ASC)}.\hskip 1em plus 0.5em minus 0.4em\relax IEEE, 2017, pp. 1532--1536.

\bibitem{liu2020speech}
Z.-T. Liu, A.~Rehman, M.~Wu, W.-H. Cao, and M.~Hao, ``Speech personality
  recognition based on annotation classification using log-likelihood distance
  and extraction of essential audio features,'' \emph{IEEE Transactions on
  Multimedia}, vol.~23, pp. 3414--3426, 2020.

\bibitem{kopru2020multimodal}
B.~K{\"o}pr{\"u} and E.~Erzin, ``Multimodal continuous emotion recognition
  using deep multi-task learning with correlation loss,'' \emph{arXiv preprint
  arXiv:2011.00876}, 2020.

\end{thebibliography}

\end{document}